\documentclass[fleqn,twoside]{article}
\usepackage{espcrc2}
\usepackage{epsfig}

\usepackage{graphicx}
\usepackage[figuresright]{rotating}
\newcommand{\AmS}{{\protect\the\textfont2
  A\kern-.1667em\lower.5ex\hbox{M}\kern-.125emS}}

\newcommand{\lsim}{\mathrel{\mathop{\kern 0pt \rlap
  {\raise.2ex\hbox{$<$}}}
  \lower.9ex\hbox{\kern-.190em $\sim$}}}
\newcommand{\gsim}{\mathrel{\mathop{\kern 0pt \rlap
  {\raise.2ex\hbox{$>$}}}
  \lower.9ex\hbox{\kern-.190em $\sim$}}}

\title{The hadronic models for cosmic ray physics: the FLUKA code solutions}

\author{
G. Battistoni\address[MI]{INFN, Sezione di Milano and Universit\`a di Milano,  
        Dip. di Fisica,
        via Celoria 16, I-20133 Milano, Italy}, %
        M.V. Garzelli\addressmark[MI]\thanks{Correspon. author, 
{\it e-mail}: {\tt Maria.Garzelli@mi.infn.it}
Invited talk presented at ISVHECRI 2006, International Symposium on
Very High Energy Cosmic Rays, Weihai, China, August 15 - 22, 2006},
	E. Gadioli\addressmark[MI],
        S.~Muraro\addressmark[MI],
        P.R. Sala\addressmark[MI],
        A. Fass\`o\address[SLAC]{SLAC, Stanford, CA 94025, US},
	A. Ferrari\address[CERN]{CERN, CH-1211 Geneva 23, Switzerland},
        S.~Roesler\addressmark[CERN],
	F. Cerutti\addressmark[CERN],
	J. Ranft\address[SIE]{Siegen University, Fachbereich 7 - Physik, 
        D-57068 Siegen, Germany},
        L.S. Pinsky\address[HOU]{University of Houston, Department of
        Physics, TX 77204-5005 Houston, US},
	A. Empl\addressmark[HOU], 
        M.~Pelliccioni\address[FRA]{INFN, via Fermi 40, 
I-00044 Frascati (Rome), Italy}
	and
        R. Villari\address[ENEA]{ENEA, via Fermi 45, I-00044 Frascati (Rome), 
Italy}
       }

\begin{document}

\begin{abstract}
FLUKA is a general purpose Monte Carlo transport and interaction code used 
for fundamental physics and for a wide range of applications. 
These include Cosmic Ray Physics (muons, neutrinos, EAS, 
underground physics),  both for basic research 
and applied studies in space and atmospheric flight dosimetry
and radiation damage. A review of the hadronic models available in FLUKA
and relevant for the description of cosmic ray air showers is presented
in this paper. 
Recent updates concerning these models are discussed. The FLUKA
capabilities in the simulation of the formation and propagation of EM and 
hadronic showers in the Earth's atmosphere are shown. 
\vspace{1pc}
\end{abstract}

\maketitle
\section{Introduction}

Extended Air Showers (EAS)
are originated by highly energetic cosmic rays, which   
interact with air atoms in the Earth's atmosphere, 
producing elementary hadrons
and nuclear fragments. Neutral pions immediately decay in two photons,
which in turn interact with other particles, ge\-ne\-ra\-ting  
electromagnetic cascades, while charged
pions and kaons, as well as other hadrons, interact and/or decay, depending
on their energy and air density, leading
to the hadronic component of the shower. 
Thus, to understand a process as complex as an air shower, 
models and codes capable of describing the evolution of 
both the electromagnetic and the hadronic component are needed. 
Monte Carlo codes have been developed in the last few years for this
purpose, such as the CORSIKA simulation package~\cite{corsika}, 
allowing to choose
among different models and model combinations for the description of
EAS formation and propagation. 
%In particular, among the high energy hadronic generator 
%implemented in CORSIKA, we mention old and new versions 
%of  SIBYLL, DPMJET, QGSJET, NEXUS and EPOS,
%while, as low energy models, i.e. for energies below 100 - 200 GeV, 
%GHEISA and FLUKA are mostly applied.  

In this paper, 
the FLUKA Monte Carlo code~\cite{flukacern,chep03} is applied
to the study of EAS induced by
cosmic rays with primary energies up to 1000~TeV. 
These results rely entirely on the use of the FLUKA
code for transport, interactions and decays
of leptons and $\gamma$s at all energies and 
for hadron-nucleus interactions 
at energies below 20~TeV, while the DPMJET code~\cite{dpmjet,dpmranftold}
is used for all nucleus-nucleus interactions and for hadron-nucleus
interactions at the highest energies. 
Our simulations are completely independent with respect
to CORSIKA, which contains only a partial version of FLUKA, for use
at low energies only (E $\lsim$ 100--200~GeV). 
The energy range spanned in this work is of interest for 
cosmic ray experiments aiming at the determination of the primary 
spectra and composition for energies up to the knee region, such
as the ATIC, the RUNJOB and the KASCADE experiments.

It is worth mentioning that FLUKA has not been specifically designed
for the study of Cosmic Ray Physics, but can be applied also in this
field. At energies $<$ 30~TeV, interesting results concerning the
prediction of inclusive lepton fluxes have already been obtained.
%and are the subject of different papers. 
In particular, the first 3-dimensional 
calculation of the atmospheric neutrino flux due to Cosmic Rays~\cite{fluxnu} 
was made with FLUKA, and, more recently, 
data con\-cer\-ning muons detected by the balloon-borne
BESS spectrometer at various depths and by the L3+C spectrometer 
located at CERN have been quite succesfully reproduced~\cite{now2006}.   
\section{Validation of the FLUKA models}
In this section some aspects of the FLUKA models relevant for Cosmic
Ray description are briefly discussed, together with the validation of
the mo\-dels by means of data collected at ac\-ce\-le\-ra\-tors. Actually,
even the most recent ac\-ce\-le\-ra\-tor data allow to test theoretical 
Monte Carlo models only for some specific projectile beam - target combinations
at energies 
well below the maximum primary energies,
while, for the si\-mu\-la\-tion of EAS processes, one has to extrapolate
the mo\-dels to many more combinations and to
the highest energies, on the basis of
theoretical considerations.  
\subsection{FLUKA: a brief introduction}
\begin{figure}[tbh]
\begin{center}
\includegraphics[bb=33 40 498 517, width=0.23\textwidth]{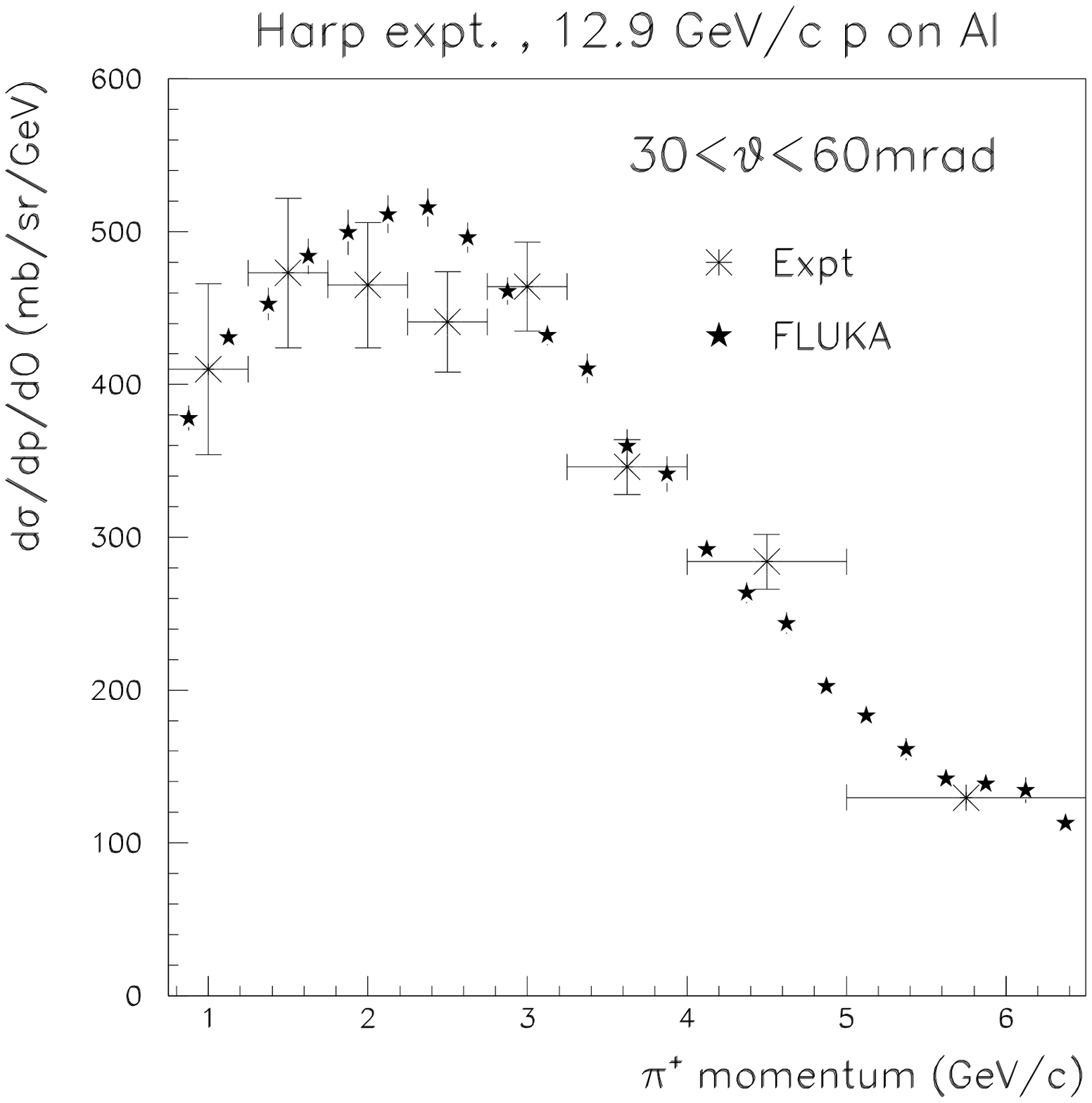}
\includegraphics[bb=33 40 498 517, width=0.23\textwidth]{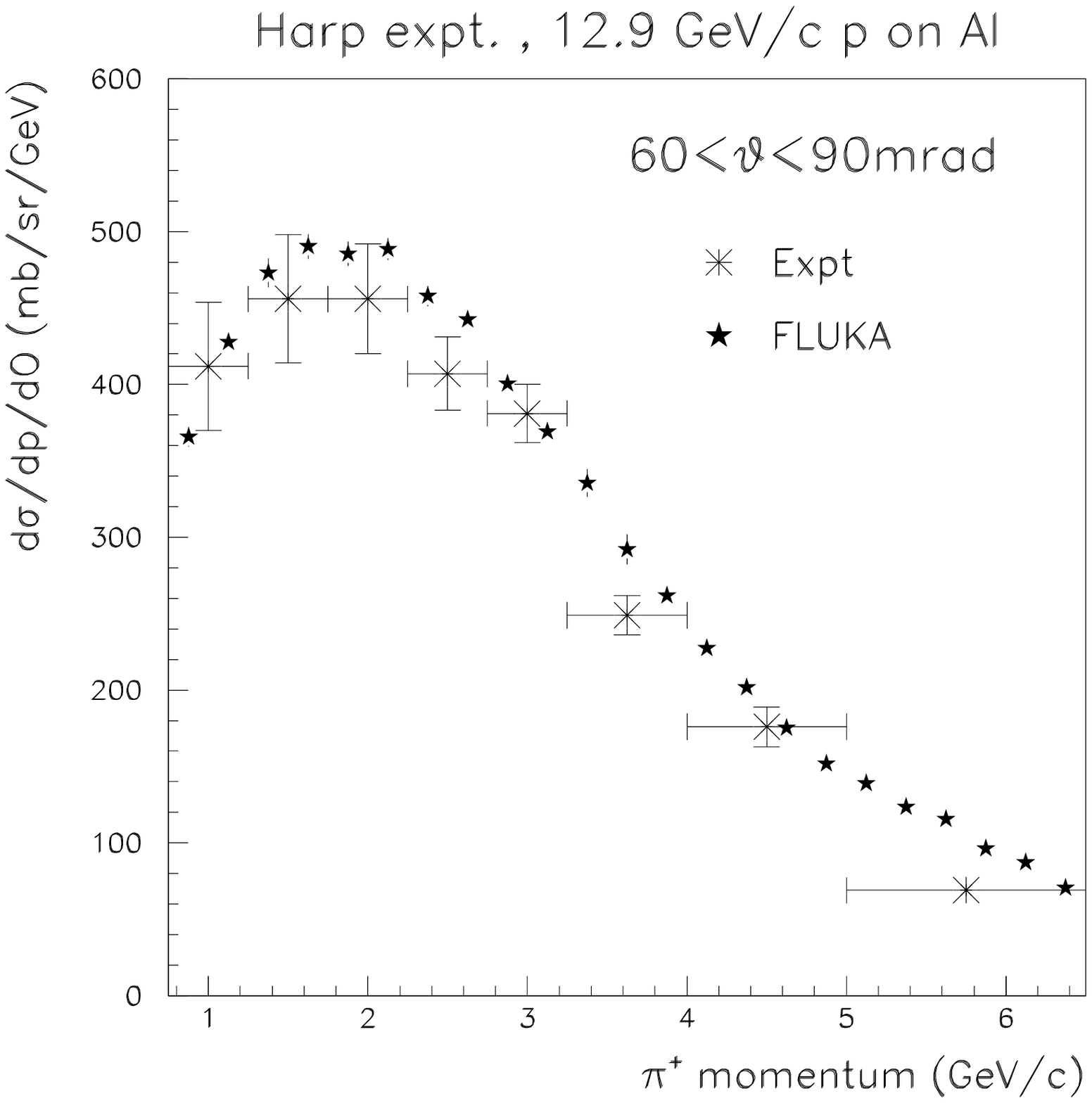}
\includegraphics[bb=33 40 498 517, width=0.23\textwidth]{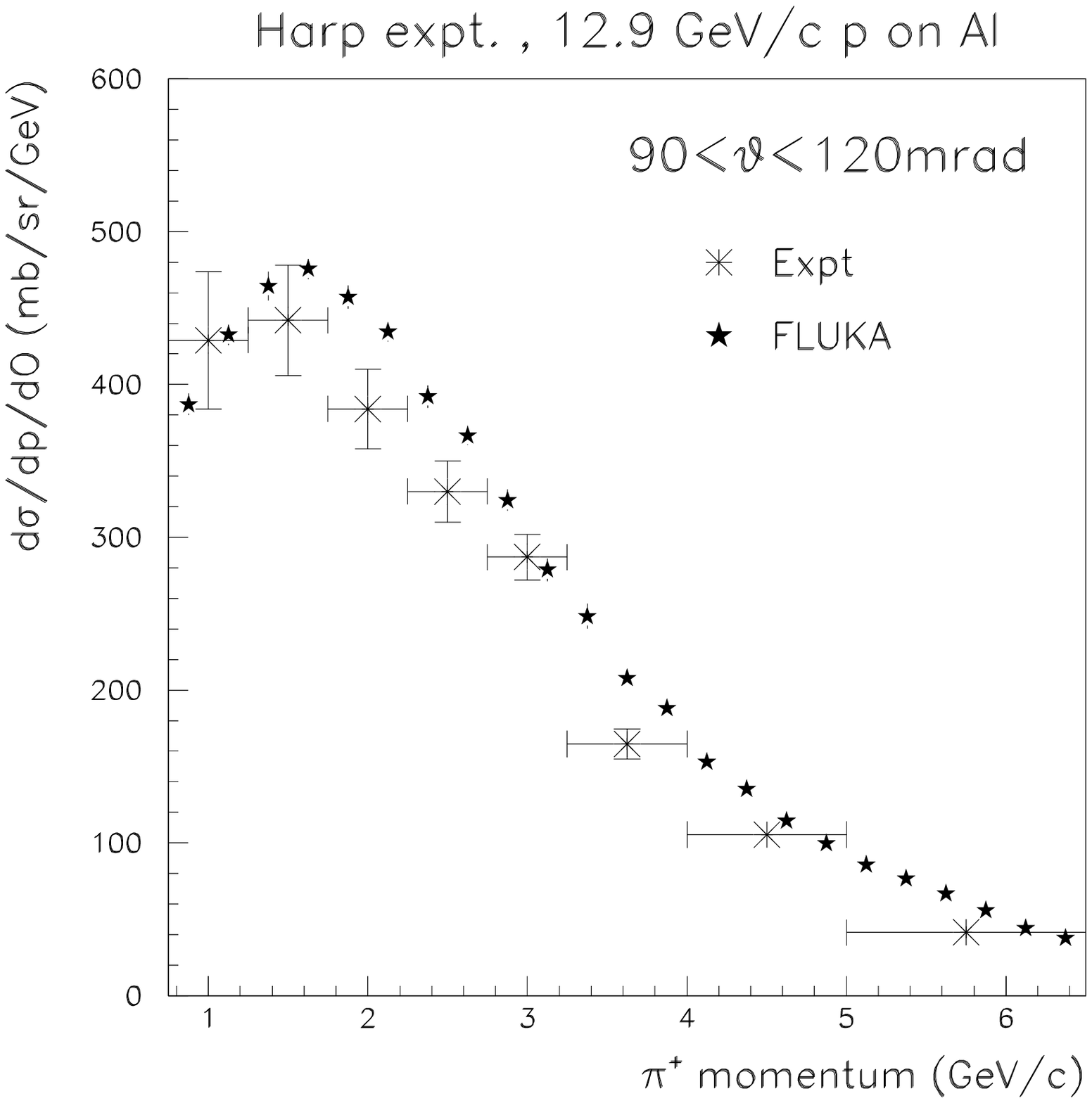}
\includegraphics[bb=33 40 498 517, width=0.23\textwidth]{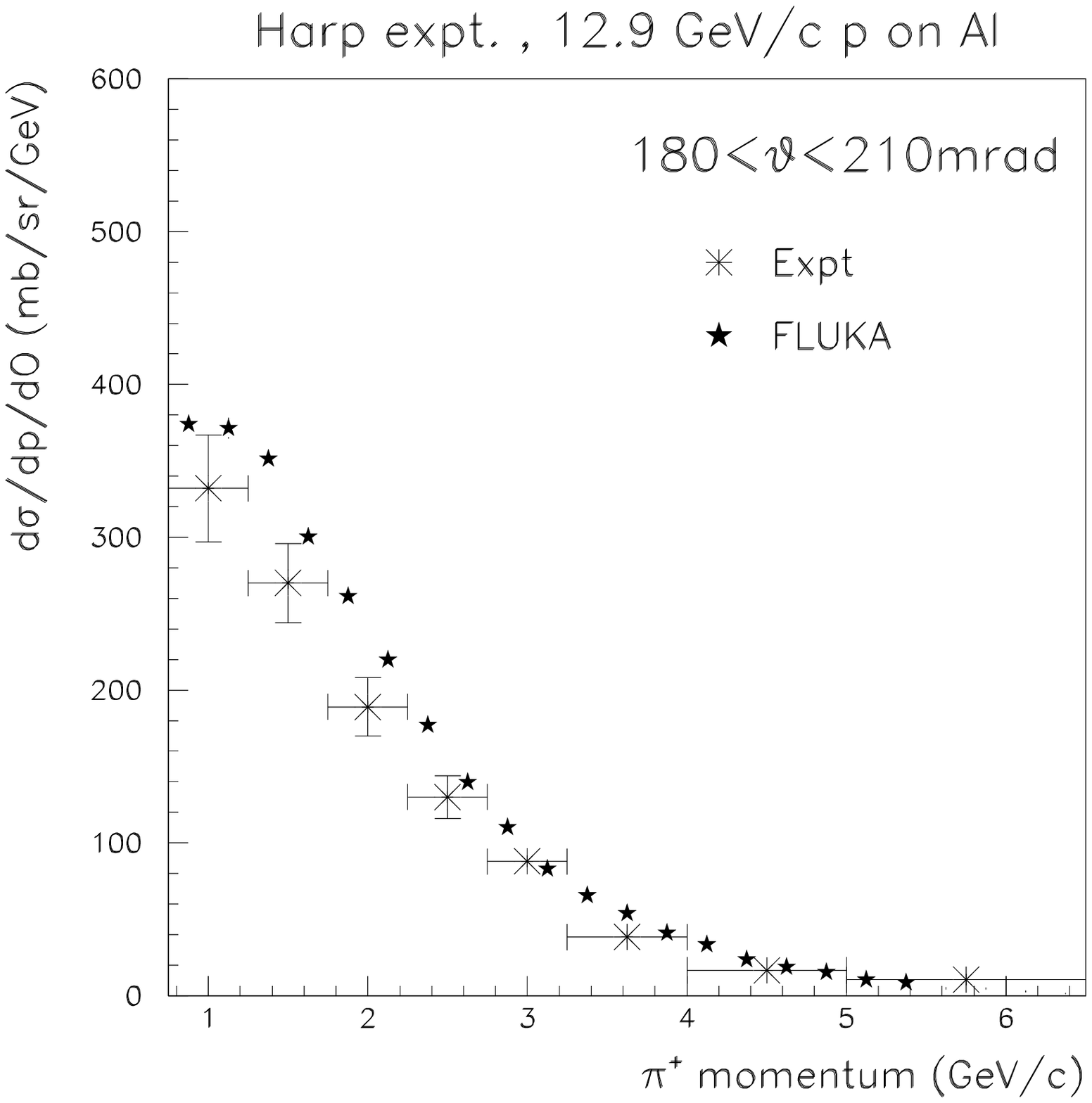}
\vspace*{-3mm}
\caption{
Computed  $\pi^+$ double 
differential production cross section for
12.9~GeV/c protons on Aluminum for different angular ranges, 
compared with HARP experimental data~\protect\cite{HARP}.
\label{fig:harp}
}
\end{center}
\vspace*{-3mm}
\end{figure}
\begin{figure}[tbh]
\includegraphics[bb=30 50 498 517, width=0.23\textwidth]{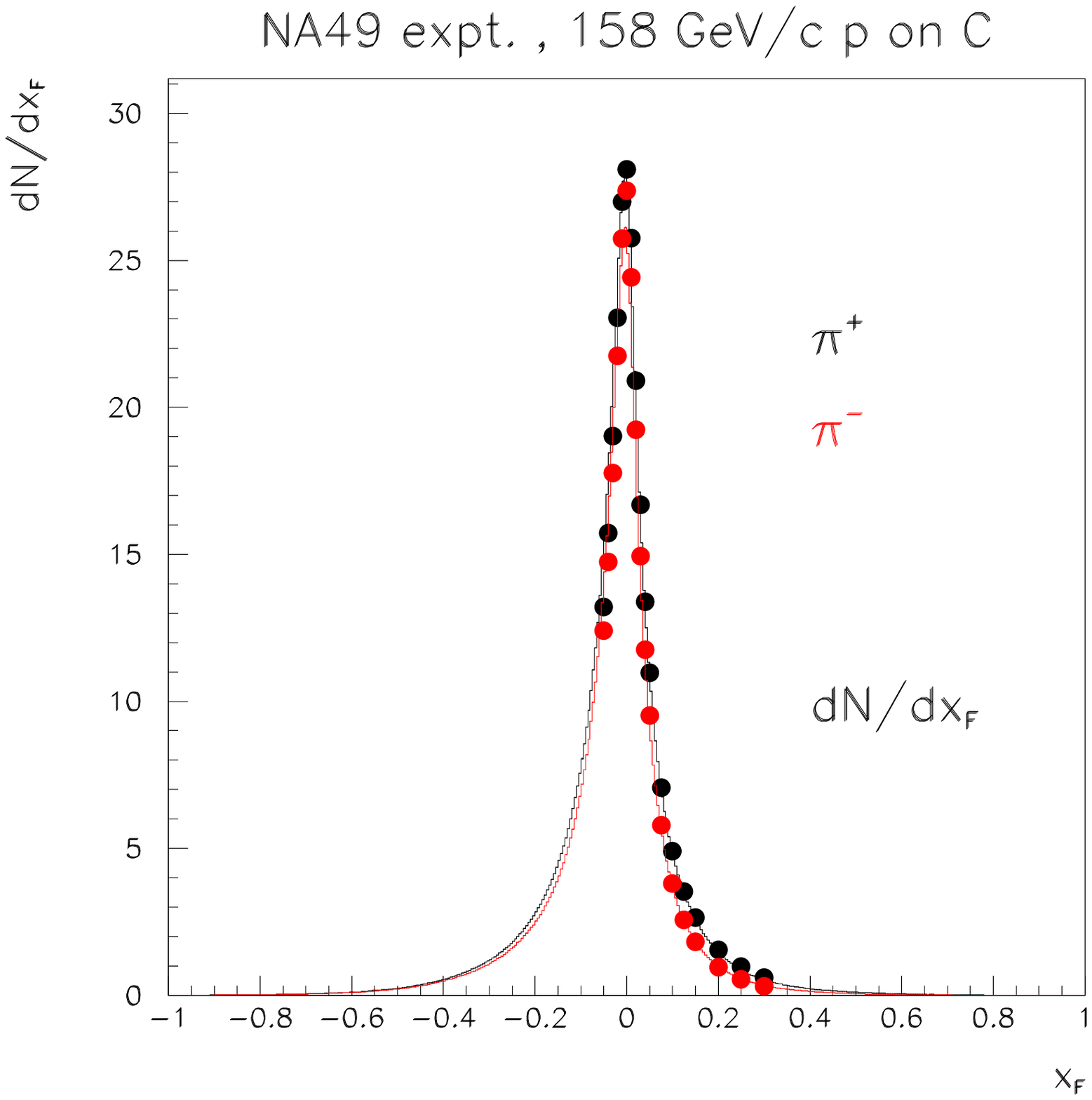}
\includegraphics[bb=30 50 498 517, width=0.23\textwidth]{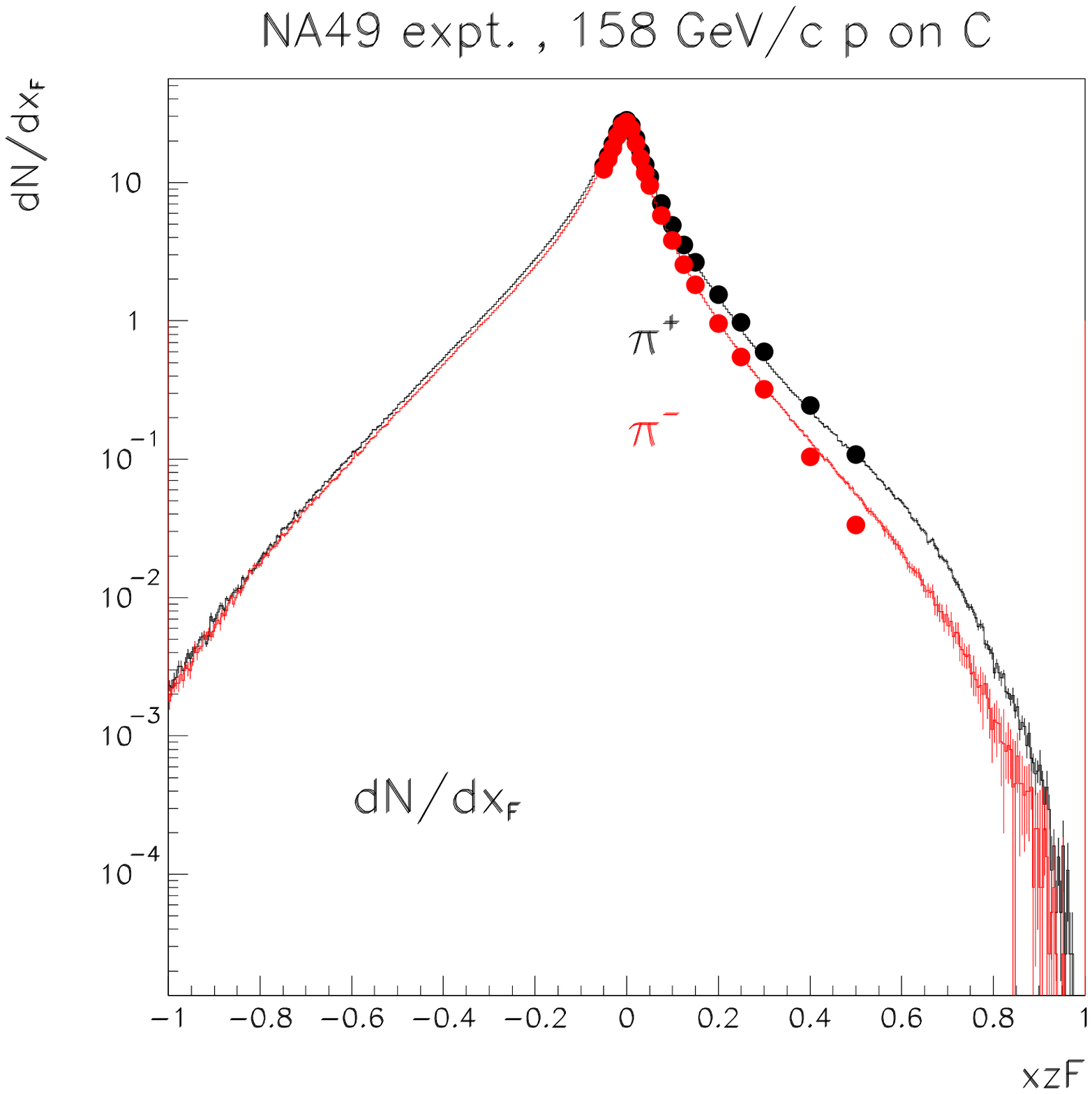}
\vspace*{-3mm}
\caption{
Feynman X distributions for $\pi^+$ and $\pi^-$ production
for proton interactions on Carbon at 158 GeV/c , 
as measured by NA49~\protect\cite{NA49} (symbols) 
and predicted by FLUKA (histograms). Linear scale on the left, 
logarithmic scale on the right.
\label{na49fig}}
\vspace*{-3mm}
\end{figure}
The modern FLUKA~\cite{flukacern,chep03} is a general purpose Monte Carlo
transport and interaction code developed since 1988 mostly by INFN and CERN
researchers, and used in many fields of physics, both for fundamental 
research and for applications. The history of the code dates back to the 
sixties when J.~Ranft developed and applied the very first version to 
accelerator shielding. The most common FLUKA 
applications are accelerator physics, calorimetry, shielding design, 
radioprotection 
and dosimetry, while new ones, such as hadrontherapy, are promisingly
growing. Since a few years, FLUKA is also applied to Cosmic Ray Physics,
mainly with focus on low energy cosmic rays, 
as already mentioned in the Introduction, and on their effects
for dosimetry~\cite{pelli,aerei} and space physics~\cite{space}. 
All these applications require both precise physics models, and packages
to build geometries, to simulate targets and ex\-pe\-ri\-mental setups. 
Both topics are the subject of constant development and improvement in FLUKA. 

As far as the geometry is concerned, a combinatorial geometry package is 
included into the code, allowing to create complex geometries, which
can be visualized by means of apposite gra\-phi\-cal tools. 
This allows to simulate
3-D particle propagation, and, as far as Cosmic Rays are concerned,
to study showers characterized by whichever inclination angle and direction 
with respect to the Earth, 
and to take into account in a precise way
the effect of different atmospheric compositions and magnetic field  
configurations. The same code can be used for the simulation of cosmic
ray detectors, including surrounding materials.

The FLUKA physical 
models are based as far as possible on
theoretical microscopic models, with the advantage, with respect 
to pa\-ra\-me\-te\-ri\-zed inclusive models, of preserving correlations, and  
of being predictive in the regions where experimental data are not
a\-vai\-la\-ble. The model parameters are fixed once, internally to the code,
for all projectile-target
combinations and energies, and cannot be modified by the user. 
The theoretical models are continuously 
benchmarked against newly available experimental data. The
code is under continuous development and made a\-vai\-la\-ble on the website
{\it http://www.fluka.org}. The development and maintenance are performed
under a INFN--CERN agreement.    
\subsection{E.M. and muon transport in FLUKA}
For historical reasons, FLUKA is best known for its hadron event 
generators, but since more than 17 years FLUKA can handle with similar or
better accuracy electromagnetic effects~\cite{MC2000em}. 
 Briefly, the energy range covered by
this sector of FLUKA is very wide: the program can transport photons
and electrons over about 12 energy decades, from 1 PeV down to 1 keV. The
e.m. part is fully coupled with the hadron sector, including the low
energy (i.e. $<$~20~MeV) neutrons. 
The simulation of the electromagnetic ca\-scade in FLUKA is very
accurate, including the Landau-Pomeranchuk-Migdal effect and a special
treatment of the tip of the bremsstrahlung spectrum. Electron
pairs and bremsstrahlung are sampled from the proper double
differential energy-angular distributions improving the common
practice of using average angles. In a similar way, the
three-dimensional shape of the hadronic cascades is reproduced in
detail by a rigorous sampling of correlated energy and angles in decay,
scattering, and multiple Coulomb scattering.
 
Bremsstrahlung and direct pair production by muons are modelled according
to state-of-the-art theoretical description and have been checked against
experimental data~\cite{Calor96,Nima394}. Muon photonuclear interactions
are also modelled.
\subsection{The FLUKA hadronic models}
\begin{figure}[tbh]
\begin{center}
\includegraphics[bb=51 51 551 770, angle=270 
%,width=0.45\textwidth
,scale=0.2]{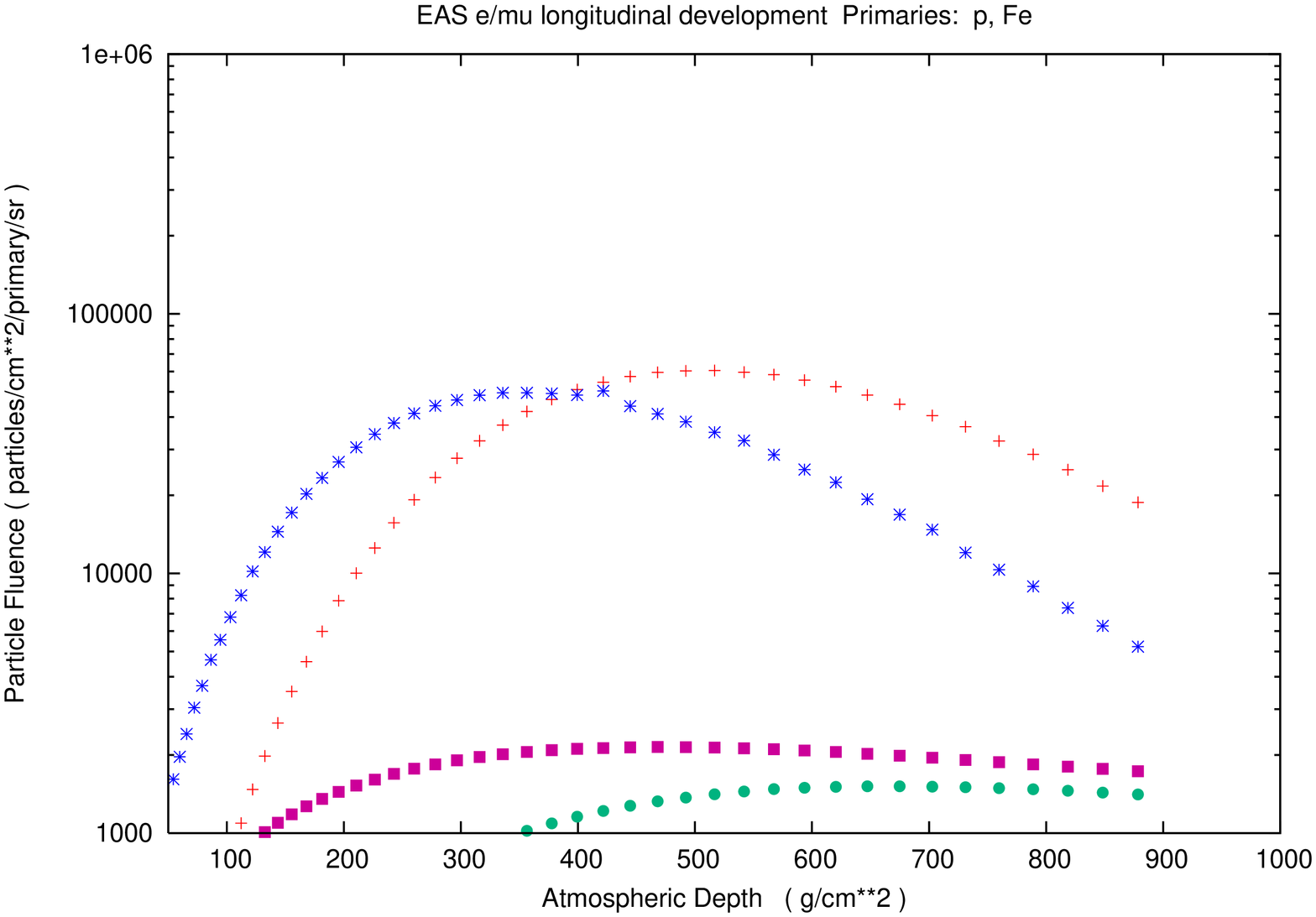}
\includegraphics[bb=51 51 551 770, angle=270 
%,width=0.45\textwidth
,scale=0.2]{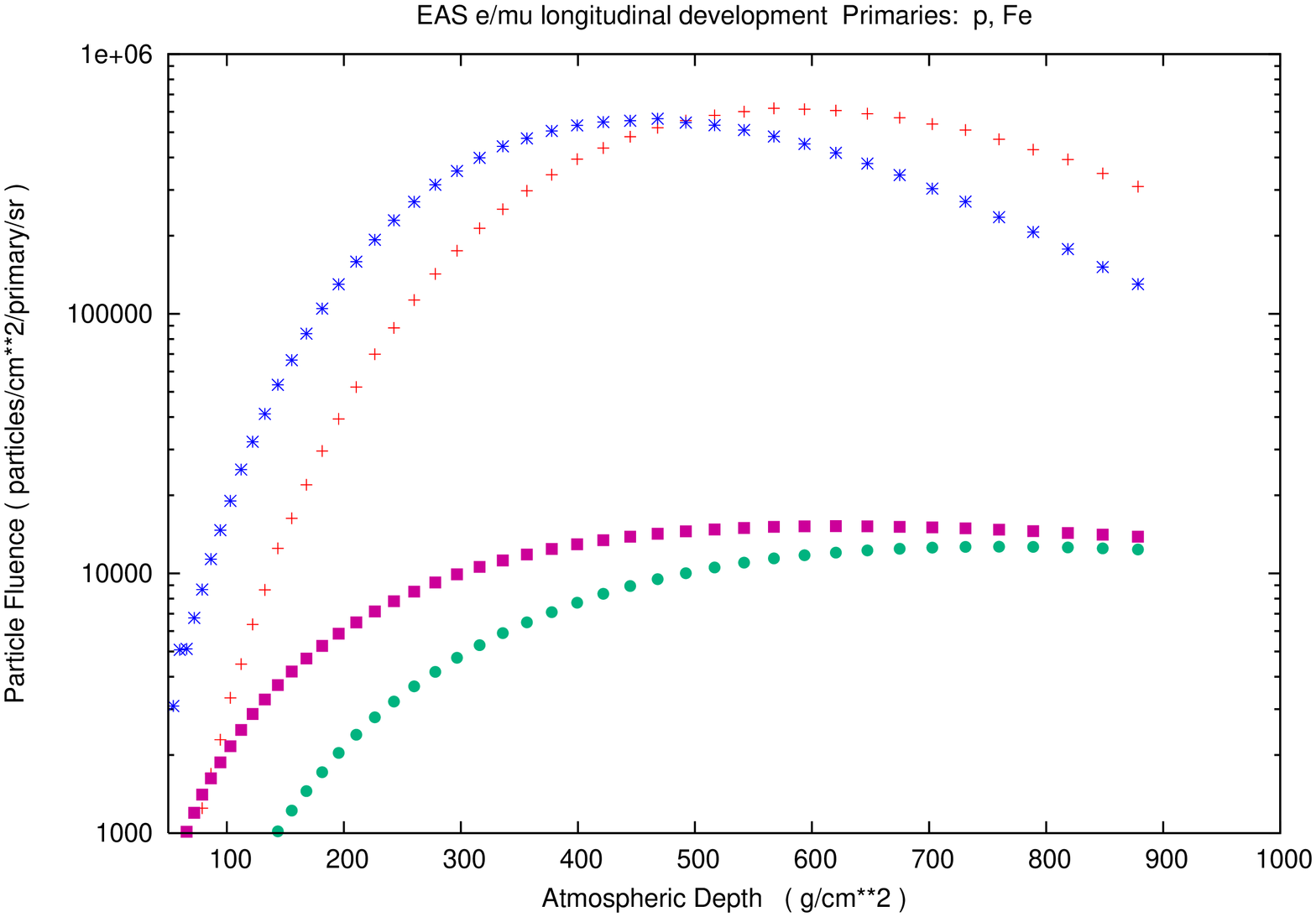}
\vspace*{-3mm}
\caption{Average e/$\mu$ fluence as a function of the atmospheric depth 
for vertical showers induced by $p$, Fe primaries with  
$10^{14}$ eV (upper panel) and $10^{15}$ eV (lower panel) energies. 
In each figure $e$ from Fe (asterisks),
$e$ from $p$ (+ symbols),  $\mu$ from Fe (filled squares) and $\mu$ from $p$ (filled circles) are shown.     
The plots refer to $e$ in the energy range 1 MeV - 1 TeV
and to $\mu$ in the energy range 1 GeV - 1 TeV.   
\label{mista14}} 
\end{center}
\vspace*{-3mm}
\end{figure}
\begin{figure}[tbh]
\begin{center}
\includegraphics[bb=51 51 551 770, angle=270
%, width=0.45\textwidth
,scale=0.2]{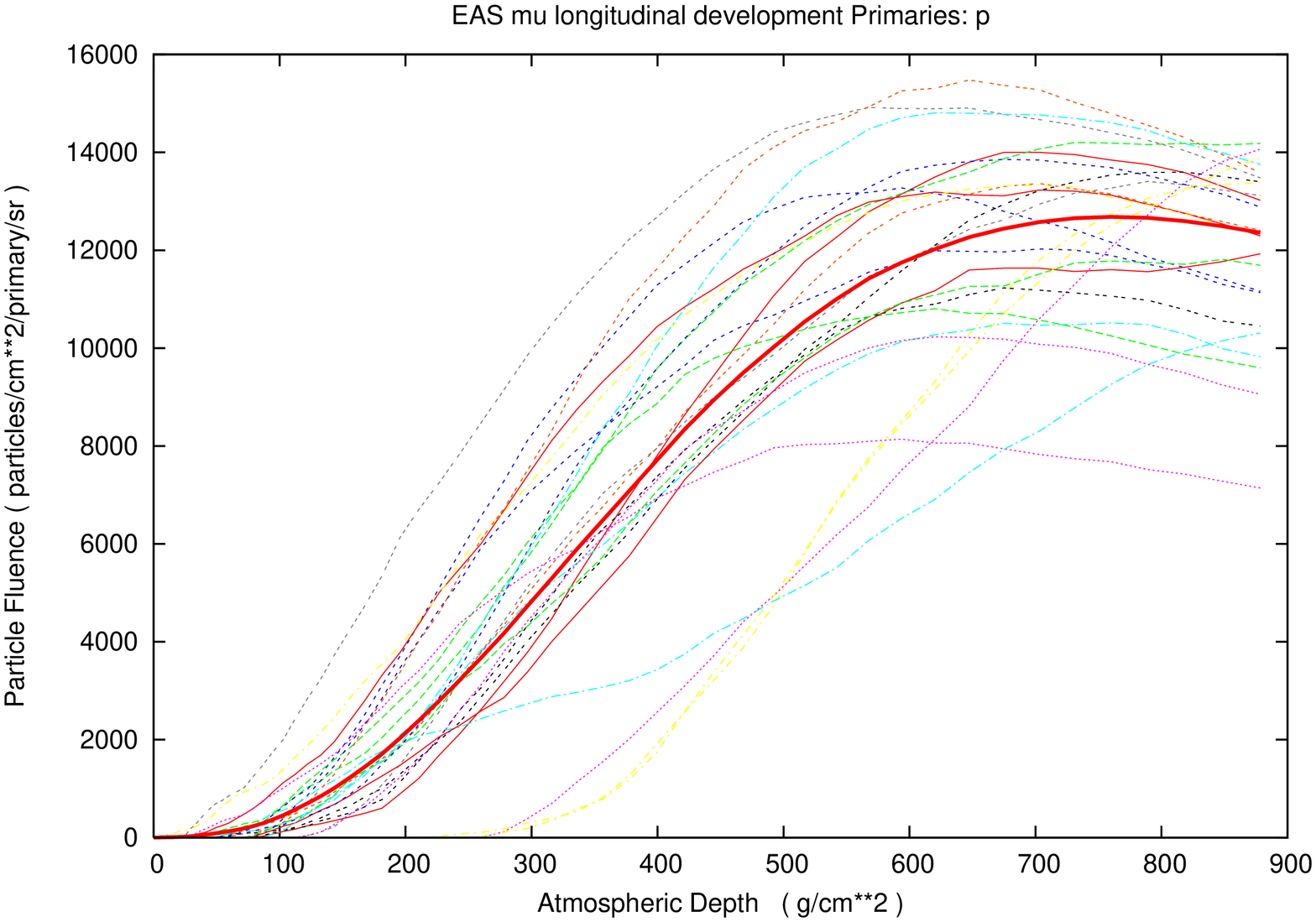}
\includegraphics[bb=51 51 551 770, angle=270 
%,width=0.45\textwidth
,scale=0.2]{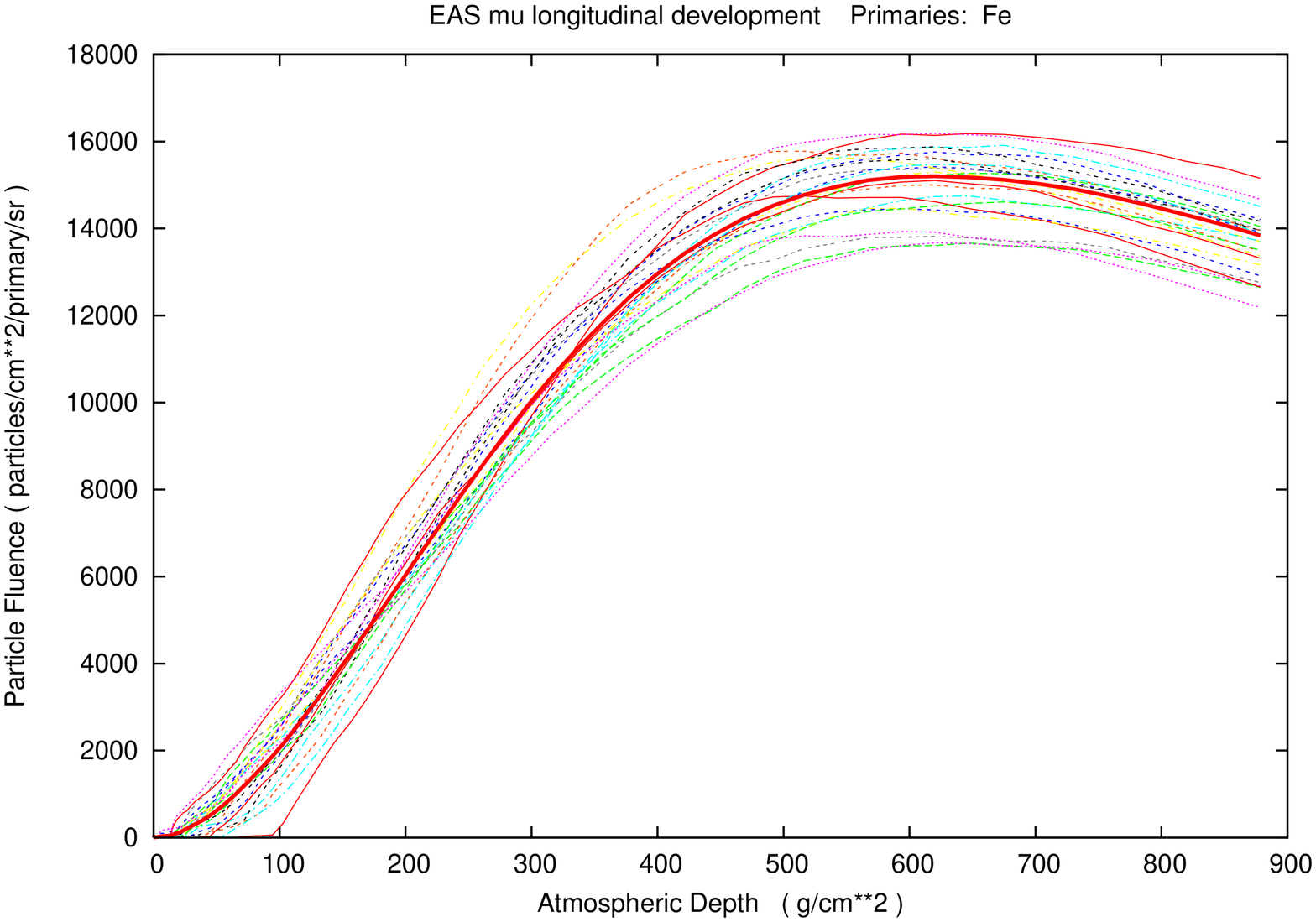}
\vspace*{-3mm}
\caption{$\mu$ fluence as a function of the atmospheric depth 
for vertical showers induced by $p$ (upper panel) and Fe (lower panel)
primaries with $10^{15}$ eV  energy. 
Each line in each figure is the result for a different shower. 
In case of $p$ showers, the development of the hadronic component 
shows larger fluctuations,
also related to the depth of the first interaction. 
See also Fig.~\ref{mista14}.       
\label{muvariance}}
\end{center}
\vspace*{-3mm}
\end{figure}
\begin{figure}[tbh]
\begin{center}
\includegraphics[bb=100 51 530 770, angle=270
%, width=0.45\textwidth
,scale=0.25]{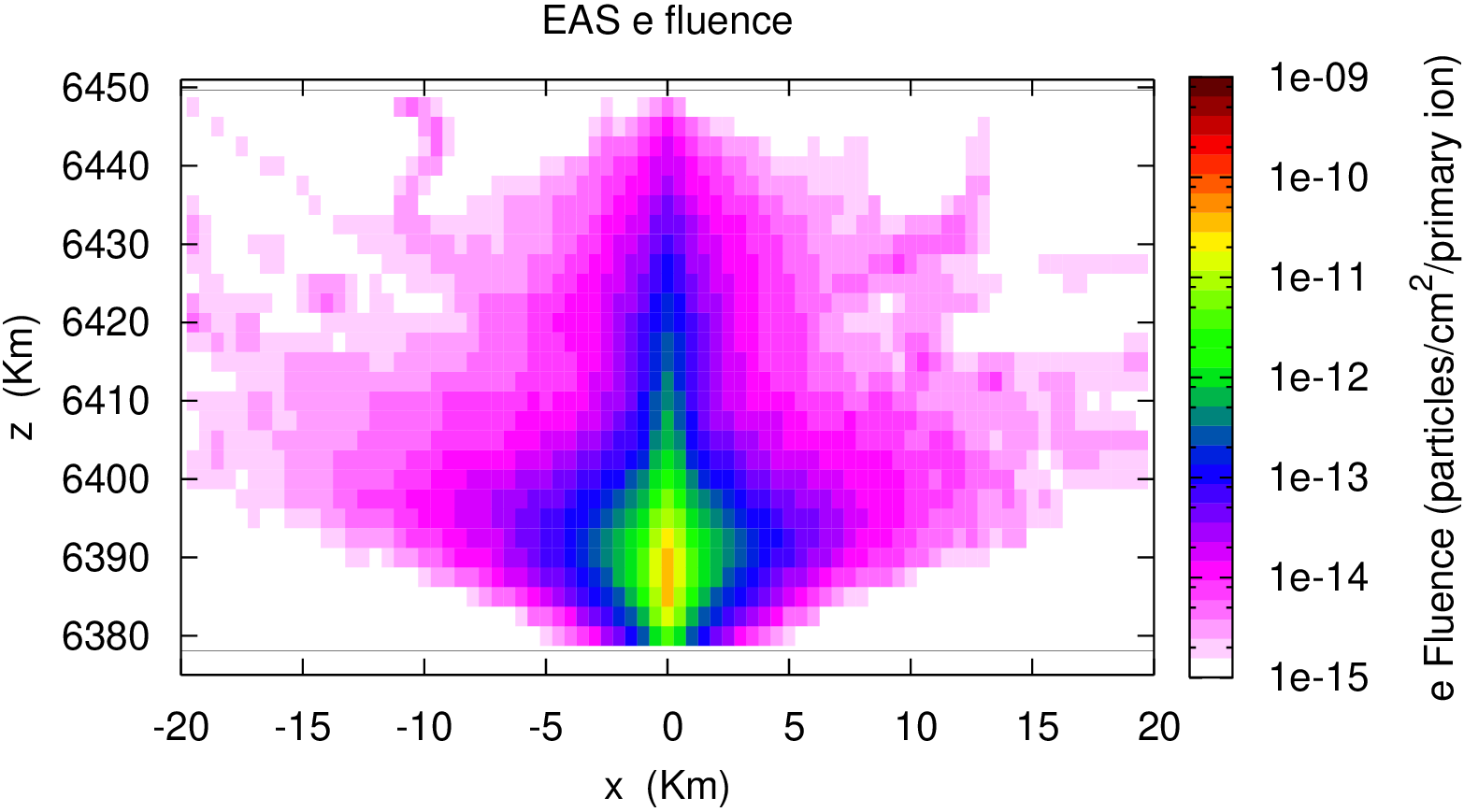}
\includegraphics[bb=100 51 530 770, angle=270
%, width=0.45\textwidth
,scale=0.25]{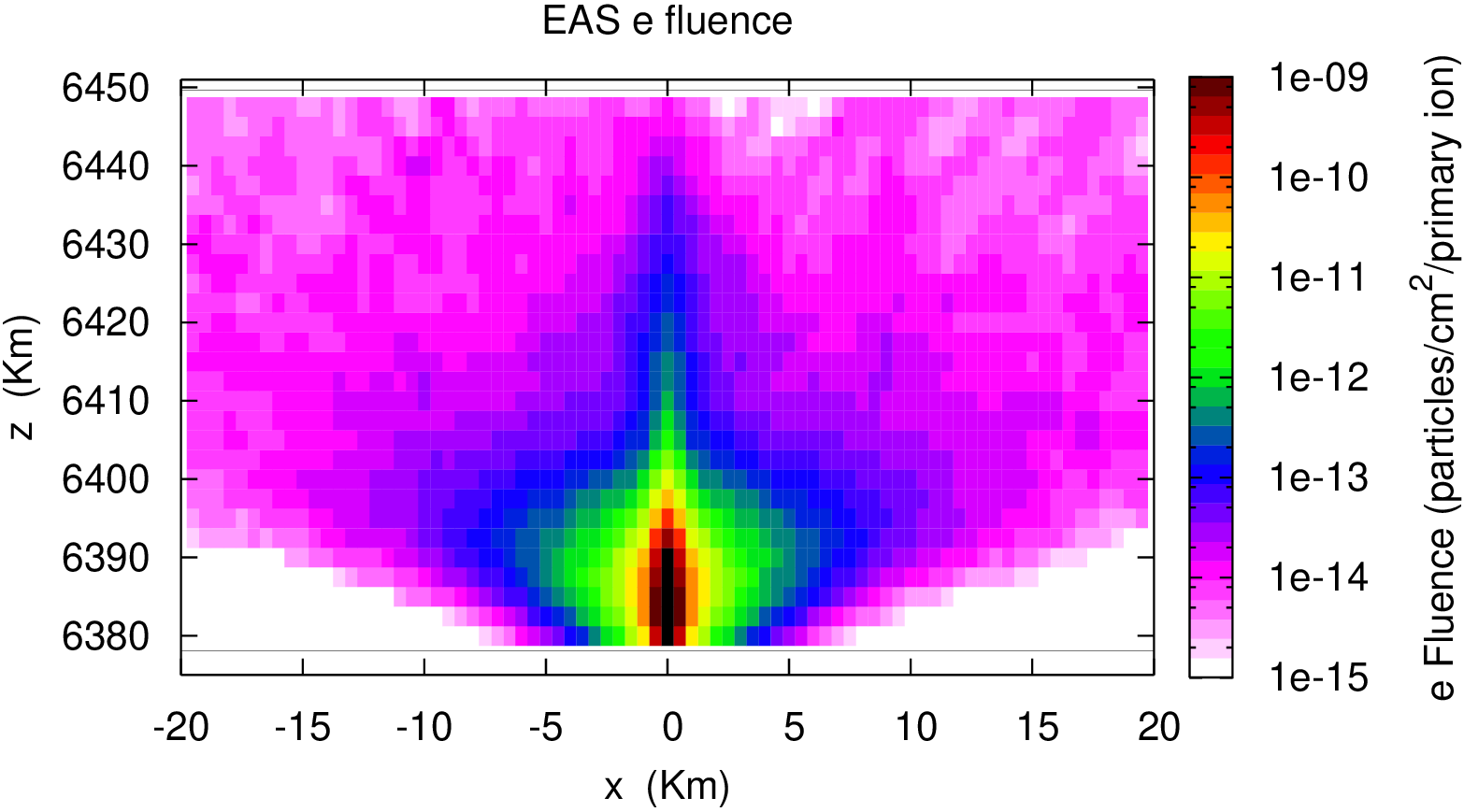}
\vspace*{-3mm}
\caption{Spatial distribution of $e$ fluence 
(particles/$\mathrm{cm}^2$/primary) 
for vertical showers induced by Fe 
primaries with $10^{13}$ eV (upper panel)
and $10^{15}$ eV (lower panel)  energies. The results are
obtained as an average on $\sim$ 100 events for each energy. The primaries
come from the top of the atmosphere (top of
each figure) and pro\-pa\-ga\-te towards the Earth's surface
located at $\sim$ 6378 km with respect to the Earth's center. 
\label{elec}}
\end{center}
\vspace*{-3mm}
\end{figure}
\begin{figure}[tbh]
\begin{center}
\includegraphics[bb=51 51 551 770, angle=270
%, width=0.45\textwidth
,scale=0.25]{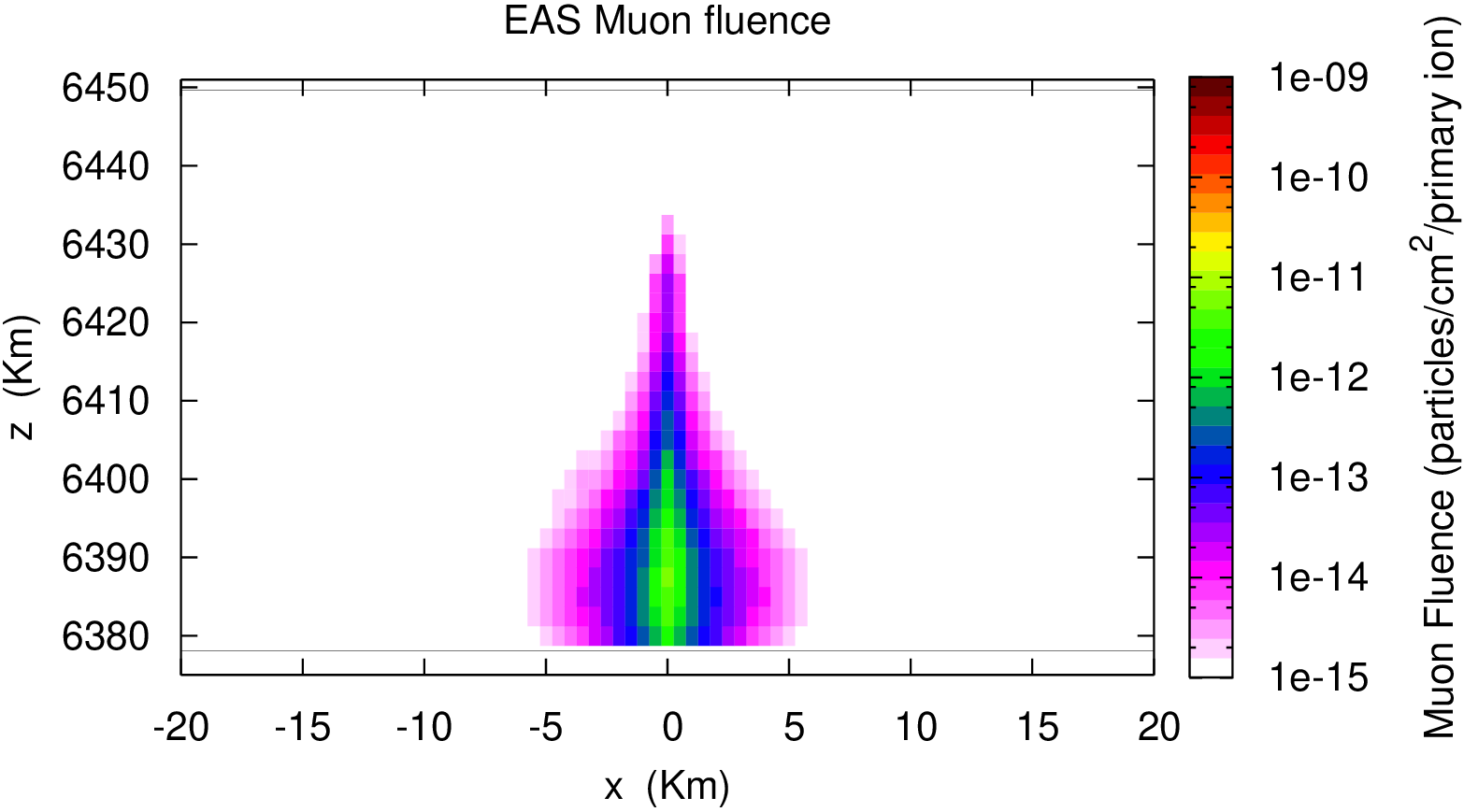}
\includegraphics[bb=51 51 551 770, angle=270
%, width=0.45\textwidth
,scale=0.25]{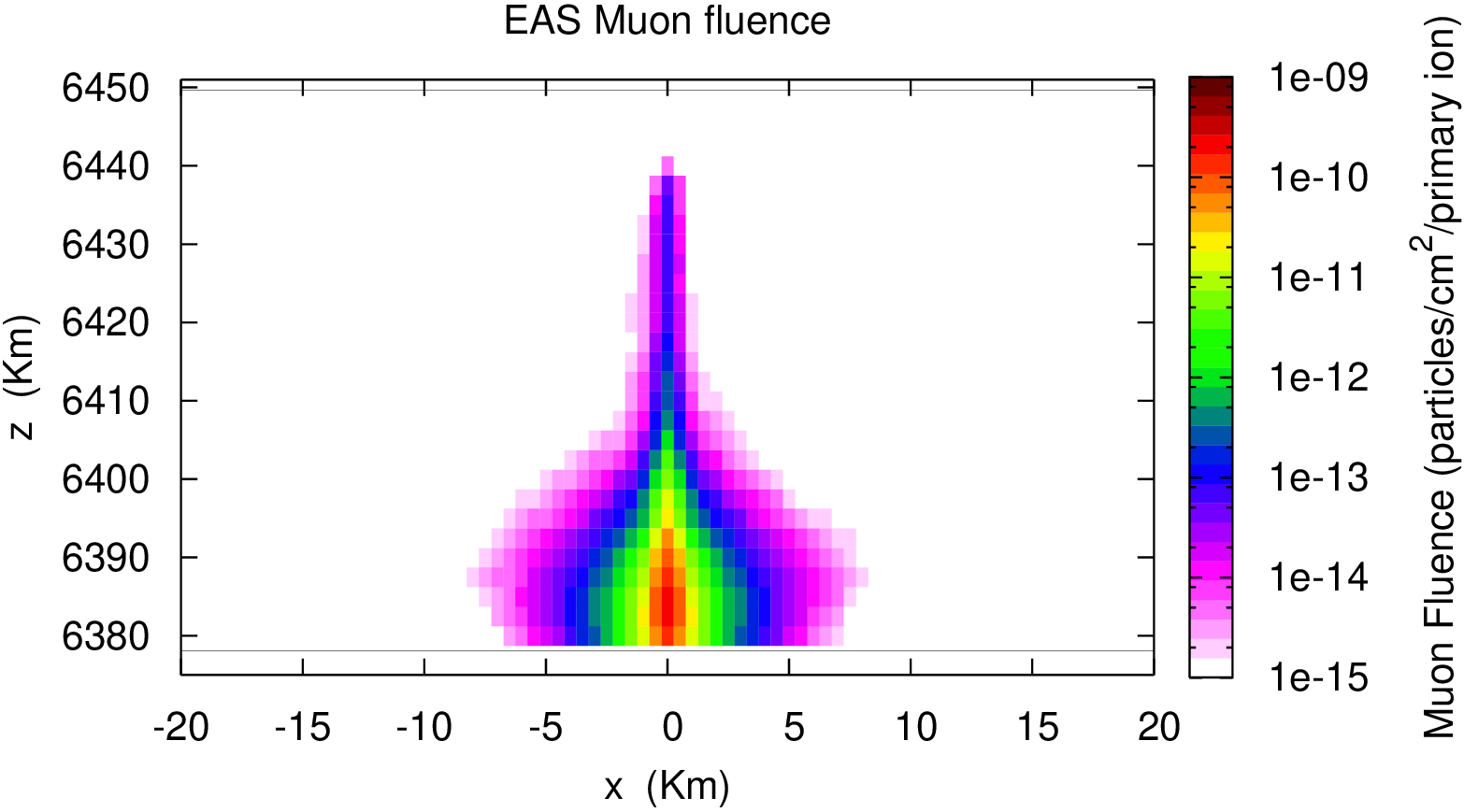}
\vspace*{-3mm}
\caption{Same as Fig.~\ref{elec} for 
$\mu$ fluence (particles/$\mathrm{cm}^2$/primary) in
vertical showers induced by Fe 
primaries with $10^{13}$ eV (upper panel)
and $10^{15}$ eV (lower panel)  energies.
\label{muons}}
\end{center}
\vspace*{-3mm}
\end{figure}
A basic description of hadronic interactions in FLUKA can be found 
in~\cite{Trieste}.
Hadron-nucleon interactions at energies below a few GeV 
are si\-mu\-la\-ted in FLUKA by the isobar model, through resonance
production and decay, and by taking into
account elastic, charge and strangeness exchange. 
Elementary hadron-hadron collisions at energies above a few GeV are described
thanks to an implementation of the Dual Parton Model (DPM)~\cite{capella},
coupled to a hadronization scheme. The Dual Parton Model allows for the
description of soft collision processes (i.e. processes cha\-rac\-te\-rized
by low $p_T$, $p_T << Q_0$, where $Q_0$ is a momentum 
scale of the order of a few GeV),
that cannot be described by means of pQCD, due to 
fast rise of $\alpha_S$ with decreasing momentum.
DPM incorporates analiticity, duality and unitarity. 
Incoming hadrons are described by strings, which interact by 
ex\-chan\-ging closed
loop excitations, called pomerons. At low energies, the single-pomeron
exchange is the dominant contribution, while at laboratory energies 
$>>$ 1 TeV higher-order contributions, corresponding to multi-pomeron
exchange, become increasingly important. Corresponding to each pomeron
cut, built according to the recipes provided by the 
Abramoviskii-Gribov-Kancheli cutting rules,
colorless chains are built, extending from the projectile valence and sea 
quarks and antiquarks to the target ones.
In particular, the valence quarks in each 
baryon are treated as a quark-diquark pair, so that, in the
simplest case of the single-pomeron exchange amplitude for a baryon-baryon
interaction, two chains
are built, each extending from a valence quark to a valence diquark. 
Sea quarks (and the corresponding antiquarks) are ad\-di\-tio\-nal\-ly 
involved
as end-points of chains in multi-pomeron exchange amplitudes.  
%Diquark breaking mechanisms, absent in the original DPM, have recently
%been proposed, in the attempt of providing a contribution to the
%enhanced stopping experimentally observed for nuclear collisions in
%fixed target experiments.
Besides de\-ter\-mi\-ning the number of pomeron cuts and forming the chains,
the DPM allows to assign an energy and momentum to each chain, according
to the momentum distribution functions of the partons in the hadrons.
In the asymptotic regime parton masses are neglected, however, at lower
energies, finite mass effects have to be included both in the chain
building process, by suppressing the chains with invariant masses 
below the observed baryon and meson mass values and reassigning energy
and momentum to other chains to ensure energy/momentum conservation,
and in the hadronization procedure. 
Hadronization is not included
in DPM, but is performed by different models properly coupled to DPM.
At present, the hadronization scheme implemented in FLUKA is based
on an advanced version~\cite{colla} of the model by~\cite{ritter}, 
with special
emphasis to provide accuracy in the low-energy description of 
hadron formation (down to a few GeV in the lab).  

\begin{figure}[t!]
\begin{center}
\includegraphics[bb=51 51 551 770, angle=270
%, width=0.45\textwidth
,scale=0.25]{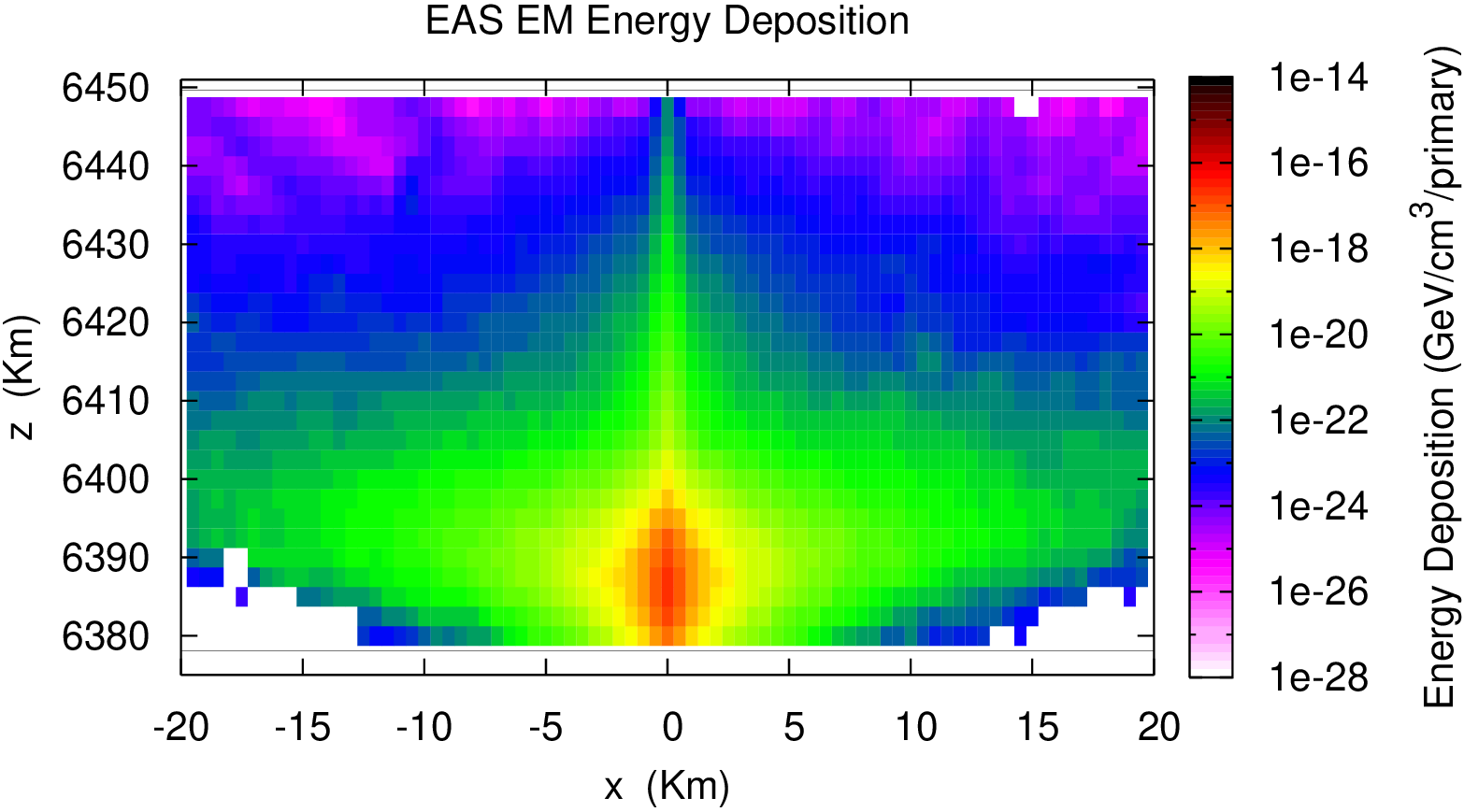}
\includegraphics[bb=51 51 551 770, angle=270
%, width=0.45\textwidth
,scale=0.25]{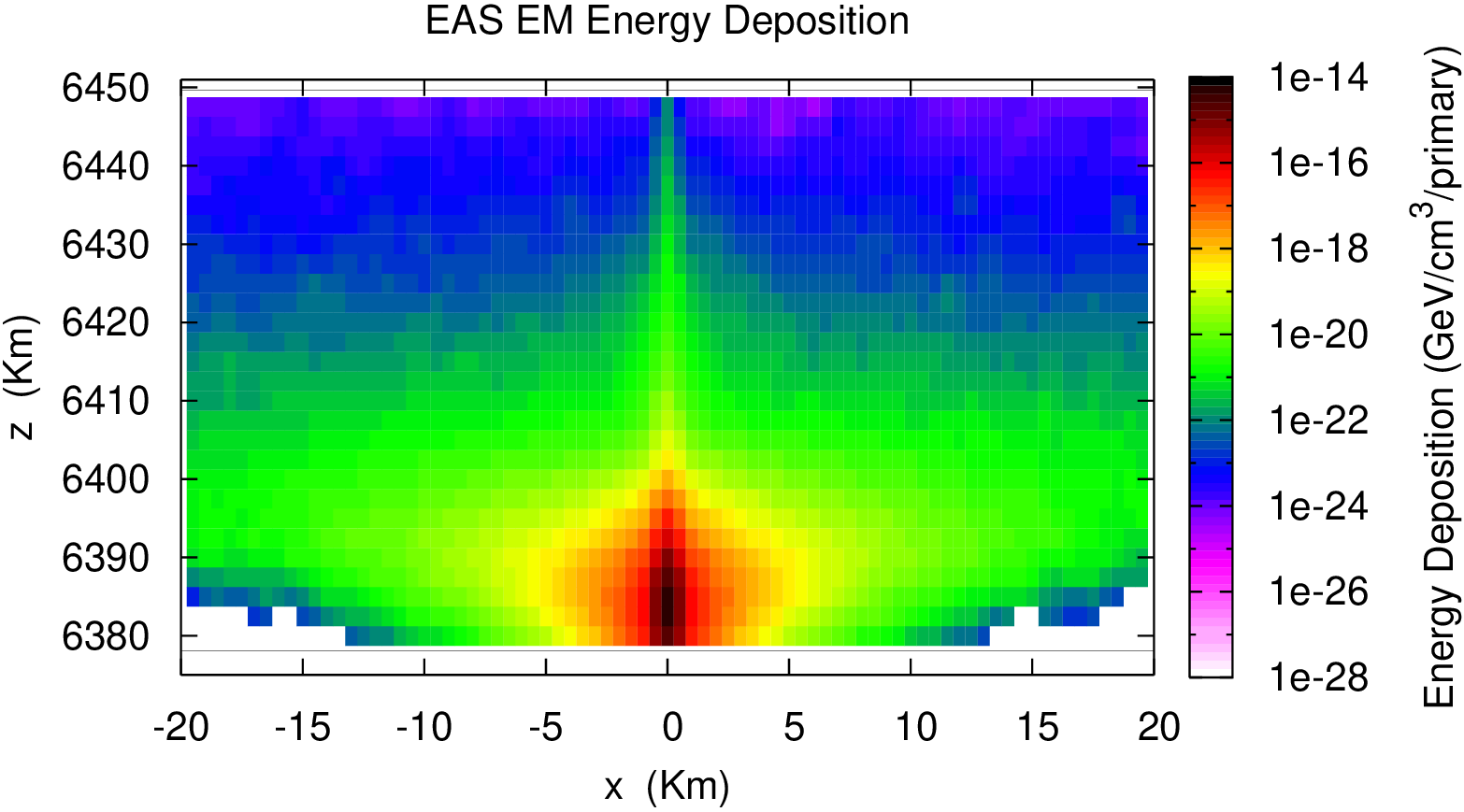}
\vspace*{-3mm}
\caption{Spatial distribution of EM energy deposition (GeV/$\mathrm{cm}^3$/primary) for vertical showers induced by Fe ion
primaries with $10^{13}$ eV (upper panel)
and $10^{15}$ eV (lower panel) incoming energies. 
See also Fig.~\ref{elec}.
\label{emene}}
\end{center}
\vspace*{-3mm}
\end{figure}
Hadron-hadron collisions are the
main building blocks of hadron-nucleus collisions. Multiple collisions
of each hadron with the nuclear constituents
are taken into account by means of the Glauber-Gribov mechanism. 
Particular efforts are devoted to the study of nuclear effects on hadron
propagation. In the last FLUKA version the
propagation in the nucleus of the hadrons resulting 
from elementary multiple collisions can be simulated
with improved accuracy with respect
to the past, by means of an
improved Gene\-ra\-lized
IntraNuclear Cascade (GINC) model, followed by pre-equilibrium and coalescence
and by de-excitation of the remaining nuclear fragments.  
All these models are included in the FLUKA submodule PEANUT, which has
been refined for several years to describe with increasing accuracy 
many low-energy processes involving nucleons, and 
has been further extended in the last few months to higher energy
events ($\mathrm{E}$ $>$ a few GeV). 
More detail about PEANUT and the issues concerning
its extension can be found in~\cite{Trieste,VarennaAlfredo}.
PEANUT is used also to simulate $\gamma$, $\nu$ and stopping $\mu$
interactions, and nucleon decay. 
Photon interactions with hadrons and muon virtual photon nuclear interactions
are simulated by means of the Vector Dominance Model.

Hadron free paths inside a nucleus can be different from those computed
on the basis of free hadron-nucleon scattering,
due to Pauli blocking, coherence length, formation and multibody processes, 
all implemented in FLUKA~\cite{Trieste,VarennaAlfredo}.
%The Pauli blocking forbids collisions which lead a nucleon in an already
%filled phase-space region, according to the Pauli exclusion principle.
In particular,
coherence length affects final states from elastic and charge-exchange scatterings: interaction products 
can not be localized better than the position uncertainty related to the
4-momentum transferred in the collision, according to the 
$\Delta x \Delta p \gsim \hbar $ relationship. Thus, reinteractions occurring
at distances shorter than the coherence length undergo interference and
can not be treated as independent.
On the other hand,
formation zone affects the reinteraction pro\-ba\-bi\-li\-ty 
of secondaries
emerging from high energy interactions.
It is supported by experimental results, which indicate that
high-energy secondary particle reinteractions inside nuclei are
strongly suppressed. 
Taking into account that a typical time for
strong interactions is $\sim$ 1 fm/c, a particle emerging from an
interaction requires some time to materialize, according to the uncertainty
principle. 

Examples of performance of the FLUKA hadronic model, 
in reproducing experimental data recently
obtained 
by the HARP and NA49 Collaborations are shown in Fig.~\ref{fig:harp}
and Fig.~\ref{na49fig}.
\subsection{The FLUKA-DPMJET interface}
\begin{figure}[t!]
\begin{center}
\includegraphics[bb=51 51 551 770, angle=270 
%,width=0.45\textwidth
,scale=0.21]{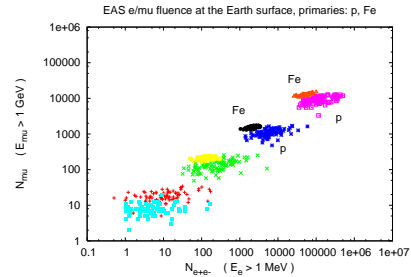}
\vspace*{-4mm}
\caption{($N_e$, $N_\mu$) correlation at the Earth sea level: dependence on the
primary mass and energy. 
Each symbol corresponds to a different shower event.
Blobs from the lower left corner to the upper right corner
refer to events originated from vertical $p$ and Fe ions with
initial energies $10^{12}$, $10^{13}$, $10^{14}$ and 
$10^{15}$ eV, respectively.  
\label{nemu}}
\vspace*{-3mm}
\end{center}
\end{figure}
\begin{figure}[t!]
\begin{center}
\includegraphics[bb=51 51 551 770, angle=270
%, width=0.45\textwidth
,scale=0.21]{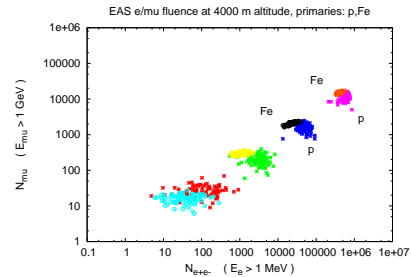}
\vspace*{-4mm}
\caption{Same as  Fig.~\ref{nemu} at a 4000 m a.s.l. altitude level. 
Calculations at such high altitudes 
are of interest for EAS experiments located at mountain sites.  
\label{highnemu}}
\end{center}
\vspace*{-3mm}
\end{figure}
Both hadronic physics at energies above 20 TeV and heavy-ion collisions
above 5 GeV have been simulated in this work thanks to the interface
of FLUKA with the DPMJET code. DPMJET-II.53 and DPMJET-III libraries 
can be both used within FLUKA.
DPMJET is also available
in CORSIKA as a possible choice among multiple high-energy event generators, 
and is widely used in the Cosmic Ray Physics community.
DPMJET has been extensively tested against data from the SPS collider, the
Tevatron and RHIC. 

Soft physics is described
in DPMJET~\cite{dpmjet,dpmranftold} 
thanks to the DPM, already introduced in the previous section.
Furthermore, hard physics processes are implemented, based on 
leading order pQCD.
Soft and hard processes contribute together
to the inelastic cross-section
according to the eikonal approximation, which provides
a unitarization scheme. The eikonal function is given
by the sum of the soft and hard components, also including 
terms from multi-pomeron exchange.  

One of the main differences between the FLUKA hadronic physics models and 
DPMJET is related to the hadronization process. In DPMJET this process
is performed by means of the JETSET chain fragmentation code included
in PYTHIA~\cite{sio}, with some modifications. 
In particular, while the PYTHIA code was
extensively tested for processes involving a significant hard component,
the transverse momentum distribution in soft chain decay has been
modified in the JETSET version included in DPMJET to better describe
soft chain hadronization. This way, the transverse momentum distributions
of K and $p$ predicted by DPMJET-III show an improved agreement with 
experimental data.

There are other differences in the description of final state interactions: 
DPMJET has its own GINC code, simplified with respect to the FLUKA one, and
%by its own 
does not take into account coalescence and pre-equilibrium emission.
Nuclear de-excitation is performed here 
%in all cases 
by means of the same de-excitation models~\cite{Trieste,ZPC1,ZPC2}
used in FLUKA.

Further refinements of DPMJET and of its interface to FLUKA 
are under way. As an example,
diquark breaking mechanisms, absent in the ori\-gi\-nal DPM, have recently
been proposed, in the attempt of providing a contribution to the
enhanced stopping experimentally observed for nuclear collisions in
fixed target experiments. The effects on the prediction of 
particle/antiparticle asymmetries, which can be compared with 
measured ones, are under investigation.  
\section{Theoretical EAS predictions}
To test the capabilities of our model,
we have simulated hundreds of vertical air shower events, 
for primary protons and iron nuclei, with energies
in the $10^{12} - 10^{15}$ eV range.  
The average longitudinal EAS e/$\mu$ profiles are shown in the upper and
lower panels of
Fig.~\ref{mista14} for  $10^{14}$ and $10^{15}$ eV 
primary energies, respectively. Electrons and positrons of energies $\ge$
1~MeV have been considered, together with muons of energies $\ge$ 1 GeV,
up to 1~TeV. 
As can be seen from each figure, at a fixed
energy, the $X^{e}_{max}$ depth decreases with increasing primary mass, due
to the behaviour of the $p$-air inelastic scattering cross-section, which
indeed increases with mass number. 
The same is true also for $\mu^{\pm}$, whose longitudinal
profile is however softer, i.e. has a maximum less peaked
than the $e^{\pm}$ one. 
More detail about the $\mu^{\pm}$ longitudinal development can be inferred 
from Fig.~\ref{muvariance}, where profiles relative to individual showers
induced by $p$ (upper panel) and Fe (lower panel) primaries 
at $10^{15}$ eV are shown. Profile fluctuations are more evident
for lower mass primaries, also related to the location of the first
primary interaction with air, which can even vary within a range of a few km
for low mass primaries and low density atmosphere.  
The muon number reflects the 
development of the hadronic component of the shower, and is
thus a basic test of the hadronic models, since muons come from charged
pion and kaon decay. These processes become important at energies below
the critical energy, i.e. the energy
for which the meson interaction and decay lengths 
are nearly the same. In particular the critical energy for $\pi$ decay
is around 150 GeV, while the one for $K$ is about ten times larger.
From these considerations, it is clear that the $\mu^{\pm}$ number 
is particularly
sensitive to a good description of the hadronic physics 
processes at low energies. We emphasize that we 
refer indeed to $\mu^{\pm}$ with energy
down to 1 GeV.
On the other hand, $\mu^{\pm}$ with energies around hundreds or thousands 
GeV in the atmosphere deeply penetrate the rock, and can be selected
and detected by underground detectors. Multi-$\mu$ and $\mu$-bundle 
events have also been studied with FLUKA, 
which has provided a succesful reproduction of experimental 
data~\cite{now2006}. 

To better identify the longitudinal shape of the shower components,
detailed maps of fluences and energy deposition
have been built. In Fig.~\ref{elec} and~\ref{muons}
the fluences of $e^{\pm}$ and $\mu^{\pm}$ 
from the top of the atmosphere, down to
the Earth's surface, are shown for Fe induced vertical showers
at  $10^{13}$  and $10^{15}$ eV.
From the figures, it is evident that in the top part 
of the atmosphere 
the $e^{\pm}$ shower profile is more extended than the $\mu^{\pm}$ one, and
that some $e^{\pm}$ 
can even propagate from lower altitudes towards higher ones.
At the considered energies, 
some $e^{\pm}$ and $\mu^{\pm}$ from the same shower
can reach ground within a few km.
Anyway, the largest fluences are found for both kind of particles in 
the spatial region
located within $\sim$ 2 - 8 km above the Earth's surface, at distances
not larger than 500 - 700 m from the shower axis.  
In Fig.~\ref{emene} detailed maps of electromagnetic energy deposition
for vertical Fe showers at $10^{13}$ and $10^{15}$ eV
are presented. These maps are strictly related to the ones for $e$ 
fluence shown in Fig.~\ref{elec}, already di\-scus\-sed. $e^{\pm}$
production thresholds have been set to 1 MeV, while $\gamma$ threshold
has been set to 0.5 MeV in our simulations. 

The ($N_e$, $N_\mu$) correlations for $p$ and Fe showers at sea level and
at 4000 m a.s.l. are shown in Fig.~\ref{nemu} and Fig.~\ref{highnemu},
respectively.
%From the left to the right of each figure,  
Results for primary energies from $10^{12}$ to $10^{15}$ eV are plotted.
%showers corresponding to $10^{12}$, $10^{13}$, $10^{14}$ and $10^{15}$ eV
%primary energies can be identified. 
It appears that for lower energy
showers and higher altitudes it is more difficult to distinguish 
the primary spectrum composition on the basis of ($N_e$, $N_\mu$) 
detection. 

The features of EAS predicted by our simulations
are in a qualitative 
%reasonable 
agreement with 
those obtained by other authors.
As a further step in our investigation, we plan to compare in more detail
our theoretical results with those obtained by means of other models,
e.g. those available in the CORSIKA code.
Beyond the question of 
the comparison of hadronic interaction models, this can be helpful to identify
also possible differences coming from different e.m. or transport  
models.
\section{Acknowledgments}
This work was partly supported by the University of Milano and INFN, 
by the Italian MURST under contract COFIN 2004, 
and by Department of Energy contract DE-AC02-76SF00515.

\end{document}